\documentclass[preprint,tightenlines,showpacs,showkeys,floatfix,
nofootinbib,superscriptaddress,fleqn]{revtex4}
\usepackage{float}
\usepackage{bm}
\usepackage{graphicx}
\usepackage{color}

\makeatletter

\providecommand{\tabularnewline}{\\}
\usepackage{graphicx}
\usepackage{color}
\makeatother

\makeatother

\begin{document}
\preprint{INHA-NTG-06/2010}

\title{Anomalous tensor magnetic moments and form factors of
  the proton in the 
self-consistent chiral quark-soliton model}

\author{Tim Ledwig}

\email{ledwig@kph.uni-mainz.de}

\affiliation{Institut f\"ur Kernphysik, Universit\"at Mainz, D-55099
  Mainz, Germany}

\author{Antonio Silva}

\email{ajsilva@fe.up.pt}

\affiliation{Faculdade de Engenharia da Universidade do Porto,
  R. Dr. Roberto Frias s/n, P-4200-465 Porto, Portugal} 

\author{Hyun-Chul Kim}

\email{hchkim@inha.ac.kr}

\affiliation{Department of Physics, Inha University, Incheon 402-751, Korea}

\date{June 2010}




\begin{abstract}
We investigate the form factors of the
chiral-odd nucleon matrix element of the tensor current. In particular, 
we aim at the anomalous tensor magnetic form factors of the nucleon
within the framework of the SU(3) and SU(2) chiral quark-soliton model. We
consider $1/N_c$ rotational corrections and linear effects of SU(3)
symmetry breaking with the symmetry-conserving quantization employed.  
We first obtain the results of the anomalous tensor magnetic moments
for the up and down quarks: $\kappa_{T}^{u}=3.56$ and
$\kappa_{T}^{d}=1.83$, respectively. The strange anomalous tensor  
magnetic moment is yielded to be $\kappa_{T}^{s}=0.2\sim -0.2$, that is compatible with zero. We
also calculate the corresponding form factors $\kappa_{T}^{q}(Q^{2})$
up to a momentum transfer $Q^{2}\leq 1\,\mathrm{GeV}^{2}$ at a
renormalization scale of $0.36\,\mathrm{GeV}^{2}$.
\end{abstract}

\pacs{13.88.+e, 12.39.-x, 14.20.Dh}
\keywords{Form factor, chiral quark-soliton model, tensor charges,
anomalous tensor magnetic moment, transversity}

\maketitle




\section{Introduction}

The transversity of the nucleon is an important issue in hadron
physics~\cite{Ralston:1980pp,Jaffe:1991ra,Cortes:1991ja} (see also the
review~\cite{review,Boffi:review}), since, together with the helicity distribution, they are
the leading-twist parton distributions with which one tries to understand the spin structure of the nucleon.
 However, it is of great difficulty to measure
the transversity of the nucleon because it decouples
from deep inelastic scattering due to its chiral-odd nature.  
Only very recently, information on the transversity of the nucleon is
available from the transverse spin asymmetry $A_{TT}$ of Drell-Yan
processes in $p\overline{p}$ reactions~\cite{Efremov:2004,
Anselimo:2004,PAX:2005,Pasquini:2006} as well as the azimuthal single
spin asymmetry in semi-inclusive deep inelastic 
scattering (SIDIS) \cite{Anselmino:tensorcharge}.

The chiral-odd nucleon matrix element is parameterized by three form
factors \cite{Diehl:GPDhFlip,Hagler:FFdecomposition} which we denote  
as $H_{T}(Q^{2})$, $E_{T}(Q^{2})$ and $\tilde{H}_{T}(Q^{2})$. The
first one was studied in Ref. \cite{Ledwig:2010tu} together with the
tensor charges within the SU(3) chiral quark-soliton model
($\chi$QSM).  Beside the tensor charges~\cite{HeJi:TensorCharges}, also the
linear combination $\kappa_{T}^{q}(0)=E_{T}^{q}(0)+2\tilde{H}_{T}^{q}(0)$ was
found to be interesting. It was found that $\kappa_T^q$ is a more fundamental quantity~\cite{DiehlHaegler:ATMM}  
than $E_{T}^{q}$ and $\tilde{H}_{T}^{q}$ by themselves and describes
the transverse deformation of the transverse polarized quark
distribution in an unpolarized nucleon. Burkardt~\cite{Burkardt:2005} 
investigated the connection of the quantity $\kappa_{T}^{q}$ to the
Boer-Mulders function $h_{1}^{\perp q}$ \cite{BoerMulders:1998} and
presented the relation   
\begin{equation}
h_{1}^{\perp q}\sim-\kappa_{T}^{q}\,\,\,,
\end{equation}
which is similar to the relation of the Sivers-function 
$f_{1T}^{\perp q}$ \cite{Sivers:1991} to the anomalous magnetic
moment $\kappa^{q}$, i.e. $f_{1T}^{\perp q}\sim-\kappa^{q}$
\cite{Burkardt:2004a,Burkardt:2004b,Burkardt:2004c}. 
The latter relation is confirmed by the HERMES collaboration
\cite{HERMES:2005}. 
The Boer-Mulders function $h_{1}^{\perp q}$ represents the asymmetry 
of the transverse momentum of quarks perpendicular to the quark spin
in an unpolarized nucleon, while the Sivers function explains the
transverse momentum asymmetry of quarks in a transversely polarized
target. Both the Sivers- and Boer-Mulders functions contribute to
SIDIS processes and a measurement of $h_{1}^{\perp q}$ can provide
information on the anomalous tensor magnetic moment
$\kappa_{T}^{q}$~\cite{Burkardt:2006}. However, it is very hard to
measure chiral-odd observables. So far, only the transversity
distribution $\delta q(x)$ was extracted
\cite{Anselmino:tensorcharge}, based on a global analysis of the data 
of the azimuthal single spin asymmetry in SIDIS processes $lp^{\uparrow}\to
l\pi X$ by the Belle~\cite{Belle:2006},
HERMES~\cite{HERMES:2005,HERMES:2006} and COMPASS~\cite{COMPASS:2007} 
collaborations. The corresponding tensor charges were obtained as
$\delta u=0.54_{-0.22}^{+0.09}$ and $\delta d=-0.23_{-0.16}^{+0.09}$
at a scale of $\mu^{2}=0.8\,\,\mbox{GeV}^{2}$  
\cite{Anselmino:2009}.

In the present work, we aim at investigating the anomalous tensor
magnetic form factors of the nucleon  and extend  our previous study on
the nucleon tensor form factors~\cite{Ledwig:2010tu}. 
The anomalous tensor magnetic moment is interpreted as a quantity
that describes how the average position of quarks with spin in the
$x$-direction is shifted to the $y$-direction for an unpolarized
target relative to the transverse center of the
momentum~\cite{Burkardt:2005}. Some amount of theoretical works on
the anomalous tensor magnetic moment was already performed, for example, 
in light-cone constituent quark models \cite{Paspquini:2005,Paspquini:2007}, on the
lattice \cite{Hagler:2008}, in the SU(2)
$\chi$QSM~\cite{Wakamatsu:2008} and estimations given  in
Ref.~\cite{Burkardt:2008}. Presently all the results indicate positive  
values of the anomalous tensor magnetic moment for the up and down
quarks, i.e. $\kappa_{T}^{u}>0$ and $\kappa_{T}^{d}>0$.

We will use the self-consistent SU(3) and SU(2) chiral quark-soliton model 
($\chi$QSM)~\cite{Christov:1995vm} in order to compute the anomalous
tensor magnetic moments. The $\chi$QSM is an effective chiral
model for QCD in the low-energy regime, consisting of   
constituent quarks and the pseudo-scalar mesons as the 
relevant degrees of freedom.  Originally, the $\chi$QSM was developed
based on the QCD instanton 
vacuum~\cite{Diakonov:1987ty,CQSM:Diakonovlecture} and has only a
few free parameters. These parameters can mostly be fixed to the meson 
masses and decay constants in the mesonic sector. The only free 
parameter in the baryon sector is the constituent quark mass that is fixed by
reproducing the nucleon electric charge radius. Hence, all other baryon observables are obtained
in exactly the same framework without adjusting further parameters. In the
past, the model was very successful in describing basically all baryon 
observables\cite{Christov:1995vm}.
 Furthermore, the renormalization scale for the
$\chi$QSM is naturally given by the cut-off parameter for the
regularization which is about $0.36\,\mathrm{GeV^2}$. Note that it is
implicitly related to the inverse of the size of instantons
($\overline{\rho}\approx
0.35\,\mathrm{fm}$)~\cite{Diakonov:1983hh,Diakonov:1985eg}.  It plays in
particular an important role in investigating the tensor charges and
anomalous tensor magnetic moments of the nucleon, since they depend
on the renormalization scale already at one-loop level.

We sketch the work as follows. In Section II we briefly explain how to
derive the anomalous tensor magnetic moments within the framework of
the $\chi$QSM. In Section III we present the results of this
work. Section IV is devoted to conclusion and summary. The explicit
expressions for the relevant densities can be found in the Appendix.




\section{Anomalous tensor magnetic form factors} 
We begin with the nucleon matrix element of the tensor quark current,
which is parametrized in terms of three tensor form factors   \footnote{In the notation of the
  generalized form factors of \cite{DiehlHaegler:ATMM} the given form
  factors are related by $H_{T}^{\chi}(q^2)=A_{T10}^\chi(t)$,
  $E_T^\chi(q^2)=B_{T10}^\chi(t)$ and
  $\tilde{H}_T^\chi(q^2)=\tilde{A}_{T10}^\chi(t)$.}
\begin{eqnarray}
\langle
N(p^{\prime},\,s_3')|\bar{\psi}(0)i\sigma_{\mu\nu} \lambda^\chi
\psi(0) | N(p,\,s_3)\rangle & = & \overline{u}
(p^{\prime},s_3^{\prime})\left[H_{T}^\chi(Q^{2})i  \sigma^{\mu\nu} +
  E_{T}^\chi(Q^{2}) \frac{\gamma^{\mu}q^{\nu} 
    -q^{\mu}\gamma^{\nu}}{2M}\right. \cr
&& \hspace{1.cm} \left.+\; \tilde{H}_{T}^\chi(Q^{2})
  \frac{(n^{\mu}q^{\nu} -  q^{\mu}n^{\nu})}{2M^{2}}
\right] u(p,\,s_3), 
\label{eq:TmatGeneral}   
\end{eqnarray}
where $\sigma_{\mu\nu}$ is the spin operator
$i[\gamma_\mu,\,\gamma_\nu]/2$ and $\lambda^\chi$ the Gell-Mann matrices where
we also include the unity by $\lambda^0=1$. The $\psi$ represents the quark field. The $u(p,\,s_3)$
denotes the nucleon spinor of mass $M$ with momentum $p$ and the
third component of the nucleon spin $s_3$. The momentum transfer $q$
and total momentum $n$ are defined as $q=p^{\prime}-p$ and
$n=p^{\prime}+p$ respectively.  The $Q^2$ is defined as $Q^2=-q^2$
with $Q^2>0$. In the previous work, we have already 
calculated the tensor form factor $H_T^\chi(Q^2)$ within the same
framework. We proceed in this work to the anomalous tensor magnetic form factor
$\kappa_{T}^\chi(Q^{2})=E_{T}^\chi(Q^{2})+2\tilde{H}_{T}^\chi(Q^{2})$. In
fact, it is technically rather complicated to consider the form factor
$E_T^\chi(Q^2)$ and also $\tilde{H}_{T}^\chi(Q^{2})$.
However, we will see that in the linear combination
$E_{T}^\chi(Q^{2})+2\tilde{H}_{T}^\chi(Q^{2})$ these difficulties drop out.

In order to extract $\kappa_T^\chi(Q^2)$ from Eq.~(\ref{eq:TmatGeneral}), we take the
components $\mu=0$ and $\nu=k$ of the tensor current with
$s_3^{\prime}=s_3=+1/2$ fixed and use the Breit frame. Multiplying the matrix element by $q^k$, we obtain 
\begin{eqnarray}
-\langle
N(p^{\prime},\,\frac{1}{2} )|\psi^{\dagger}\gamma^{k}\lambda^\chi
\psi|N(p,\,\frac{1}{2})\rangle  q^{k} & = & \frac{{\bm q}^2}{2M}
\left(H_{T}^\chi(Q^{2}) \;+ \; E_{T}^\chi(Q^{2}) \;+\;
  2\tilde{H}_{T}^\chi(Q^{2})\frac{E^{2}}{M^2}\right), 
\end{eqnarray} 
where $E=\sqrt{M^{2}+{\bm q}^{2}/4}$ is the on-shell energy of the
nucleon. It is convenient to define a combined form factor
$H_T^{*\chi}(Q^2)$ as 
\begin{equation} 
H_{T}^{*\chi}(Q^{2}) \;=\; \frac{2M}{{\bm
    q}^{2}}\int\frac{d\Omega}{4\pi}\langle
N_{\frac{1}{2}}(p^{\prime})|\psi^{\dagger}\gamma^{k}q^{k} \lambda^\chi 
\psi|N_{\frac{1}{2}}(p)\rangle\,\,\,,
\label{eq:DEF_H*}
\end{equation}
so that the third form factor $\tilde{H}_{T}^\chi(Q^{2})$ is expressed
in terms of the other three form factors as follows:  
\begin{eqnarray}
2\tilde{H}_{T}^\chi(Q^{2})\frac{E^{2}}{M^{2}} & = &
-H_{T}^\chi(Q^{2})-E_{T}^\chi(Q^{2})-H_{T}^{*\chi}(Q^{2})\,\,\,.
\end{eqnarray}

Inserting this in the definition $\kappa_T^\chi=E^\chi_T(0)+2\tilde{H}_T^\chi(0)$, we are
able to reexpress the flavor-decomposed anomalous tensor magnetic moments as 
\begin{equation}
\kappa_{T}^{\chi} = -H_{T}^{\chi}(0)-H_{T}^{*\chi}(0), 
\end{equation}
with $\chi=0,3,8$ and $\lambda^0=1$.

Thus, in the present work we aim at calculating Eq. (\ref{eq:DEF_H*})
in the $\chi$QSM and combining it with the results for $H_{T}^\chi(Q^2)$ in
the previous work~\cite{Ledwig:2010tu} to determine the
flavor-decomposed $\kappa_{T}^{q}(Q^2)$ with $q=u,\,d,\,s$.  
Explicitly, they are expressed in SU(3) as 
\begin{eqnarray}
\kappa_{T}^{u} & = &
\frac{1}{2}\left(\frac{2}{3}\kappa_{T}^{0}+\kappa_{T}^{3} +
  \frac{1}{\sqrt{3}}\kappa_{T}^{8}\right),\cr
\kappa_{T}^{d} & = &
\frac{1}{2}\left(\frac{2}{3}\kappa_{T}^{0} - \kappa_{T}^{3} +
  \frac{1}{\sqrt{3}} \kappa_{T}^{8}\right),\cr
\kappa_{T}^{s} & = & \frac{1}{3}\left(\kappa_{T}^{0} - \sqrt{3}
  \kappa_{T}^{8} \right),
\label{eq:flavor tensor}
\end{eqnarray}
and in SU(2) with $\tau^0=1$ as:
\begin{eqnarray}
\kappa_{T}^{u} & = &
\frac{1}{2}\left(\kappa_{T}^{0}+\kappa_{T}^{3}\right),\cr
\kappa_{T}^{d} & = & \frac{1}{2}\left(\kappa_{T}^{0} - \kappa_{T}^{3} \right).
\end{eqnarray}

In order to compare the present results for the form factors with
those of other works, it is essential to know at which renormalization
scale the calculation was carried out, since the tensor form factors
depend on the renormalization scale already at one-loop level. Hence,
we will use the following equation \cite{review,Evolution1} in order to
compare results obtained at different renormalization scales: 
\begin{eqnarray}
\kappa_{T}^{q}(\mu^{2}) & = &
\left(\frac{\alpha_{S}(\mu^{2})}{\alpha_{S}(\mu_{i}^{2})}\right)^{4/27} 
\left[1-\frac{337}{486\pi}\left(\alpha_{S}(\mu_{i}^{2}) -
    \alpha_{S}(\mu^{2})\right)\right]\kappa_{T}^{q}(\mu_i^{2})
\label{eq:evolve}\\  
\alpha_{S}^{NLO}(\mu^{2}) & = & \frac{4\pi}{9\ln(\mu^{2}/\Lambda^{2})}
\left[1-\frac{64}{81} 
  \frac{\ln\ln(\mu^{2}/\Lambda^{2})}{\ln(\mu^{2}/\Lambda^{2})}
  \right]
\end{eqnarray} 
with $\Lambda=0.248$ GeV and $N_{c}=N_{f}=3$.




\section{Anomalous tensor magnetic form factors in the Chiral
  Quark-Soliton Model} 

Having performed a tedious but straightforward
calculation following Refs. \cite{Kim:eleff,Christov:1995vm,Silva:Strange},  
we finally arrive at the complete expressions for the SU(3) form factors 
$H_T^{*\chi}(Q^2)$ for the cases $\chi=3,8$ and $\chi=0$ as follows:
\begin{eqnarray}
H_{T}^{*\chi}(Q^{2}) & = & \frac{2}{3}M\int dr\,\, r^{3}\, 
\left\{ j_{0}(|\bm Q|r)+j_{2}(|\bm Q| r)\right\} \,\,\langle
N^{\prime}|  \mathcal{H}_{T}^{*\chi}(r) |N\rangle,\label{eq:38CQSMdensity} \\
\mathcal{H}_{T}^{*\chi=3,8}(r) & = &
-D_{\chi8}^{(8)}\mathcal{Q}_{T0}(r)+\frac{1}{\sqrt{3}I_{1}}D_{\chi
  i}^{(8)}J_{i}\mathcal{X}_{T1}(r)+\frac{\sqrt{3}}{2I_{2}}D_{\chi
  a}J_{a}\,\mathcal{X}_{T2}(r)\cr 
& & -\frac{2}{\sqrt{3}}\frac{K_{1}}{I_{1}}M_{8}D_{8i}^{(8)}D_{\chi
  i}^{(8)}\,\,\mathcal{X}_{T1}(r)-\sqrt{3}\frac{K_{2}}{I_{2}}M_{8}D_{8a}^{(8)} 
D_{\chi a}^{(8)}\,\,\mathcal{X}_{T2}(r)\cr 
 &  & +\; 2\left\{ M_{1} D_{\chi8}^{(8)} + \frac{M_{8}}{\sqrt{3}}
   D_{88}^{(8)}D_{\chi8}^{(8)} \right\}
 \mathcal{M}_{T0}(r) +\frac{2}{\sqrt{3}} M_{8}D_{8i}^{(8)} D_{\chi
   i}^{(8)}\,\,\mathcal{M}_{T1} (r) \cr
&& +\; \sqrt{3}M_{8}D_{\chi
   a}^{(8)}D_{8a}^{(8)}\,\,\mathcal{M}_{T2} (r), \nonumber\\
\mathcal{H}_{T}^{*\chi=0}(r) & =&  -\sqrt{3}\mathcal{Q}_{T0}(r) +2\left\{ \sqrt{3}M_{1} +  M_{8}D_{88}^{(8)}\right\} \mathcal{M}_{T0}(r).
\end{eqnarray} 
The $j_0(|\bm Q| r)$ and $j_2(|\bm Q| r)$ denote the spherical Bessel
functions. The $D^{(8)}$ stand for the SU(3) Wigner functions. 
Since the nucleon state in the $\chi$QSM is expressed by the 
Wigner function~\cite{Blotz:1992pw}, the nucleon matrix elements are
finally written in terms of the SU(3) Clebsch-Gordan coefficients. 
The $I_i$ and $K_i$ are the moments of inertia of the soliton. 
The $M_1$ and $M_8$ designate the singlet and octet components of the
quark mass matrix: $M_1=(-\overline{m} + m_{\mathrm{s}})/3$ and
$M_8=(\overline{m} - m_{\mathrm{s}})/\sqrt{3}$ with
$\overline{m}=m_{\mathrm{u}}=m_{\mathrm{d}}$.  
The explicit expressions for the densities $\mathcal{Q}_{T0}(r),\cdots,
\mathcal{M}_{T2}(r)$ are given in the Appendix.  All terms which
are proportional to $M_1$ or $M_8$ are flavor SU(3) symmetry breaking terms
stemming from the strange quark mass which is treated perturbatively up to
first order. The flavor singlet part $\chi=0$ is derived from
Eq. (\ref{eq:38CQSMdensity}) by substituting $\lambda^{0}=1$ and
$D_{0\beta}^{(0)}\to\delta_{\beta8}\sqrt{3}$.  
 
We also have used the symmetry-conserving quantization  
\cite{SymmetryConQuantization} in order to obtain the above
expression.\\
Furthermore, we can deduce also the corresponding SU(2) iso-scalar ($i=0$) and
iso-vector ($i=3$) $\chi$QSM expressions for the form factors $H_T^{*0,3}(Q^2)$ as: 
\begin{eqnarray}
H_{T}^{*i}(Q^{2}) & = & \frac{2}{3}M\int dr\,\, r^{3}\, 
\left\{ j_{0}(|\bm Q|r)+j_{2}(|\bm Q| r)\right\} \,\,\langle
N^{\prime}|  \mathcal{H}_{T}^{*i}(r) |N\rangle,\label{eq:38CQSMdensity} \\
\mathcal{H}_{T}^{*i=0}(r) & = & -\sqrt{3} \mathcal{Q}_{T0}(r)\cr 
\mathcal{H}_{T}^{*i=3}(r) & =& -\frac{1}{2\sqrt{3}I_1} \mathcal{X}_{T1}(r)
\end{eqnarray}

We refer to Refs.~\cite{Kim:eleff,Christov:1995vm} for
a detailed description of how to solve the form factors numerically.
The parameters of the model consist of the constituent quark mass, the
current quark mass $\overline{m}$, strange quark mass
$m_{\mathrm{s}}$, and the cutoff mass $\Lambda$ of the proper-time 
regularization. They are, however, not free parameters but can be
adjusted to independent observables without ambiguity in the mesonic sector.     
For a given $M$ the regularization cut-off parameter $\Lambda$ and
the current quark mass $\overline{m}$ in the Lagrangian are 
fixed to the pion decay constant $f_{\pi}=93$ MeV and the physical
pion mass $m_{\pi}=140$\, MeV, respectively. Throughout this work the
strange current quark mass is fixed to
$m_{\mathrm{s}}=180\,\textrm{MeV}$ which approximately reproduces the
kaon mass. The only parameter left to fix in the baryonic sector is the constituent quark mass. The
experimental proton electric charge radius is best reproduced in the
$\chi$QSM with the constituent quark mass $M=420$ MeV.  Moreover, the
value of 420 MeV is known to reproduce the best fit to many baryonic
observables~\cite{Christov:1995vm}. Nevertheless, we have checked in
this work the $M$ dependence of the results for the anomalous tensor
magnetic moments.




\section{Results and Discussion}

We present now the results of the proton anomalous tensor magnetic form factors $\kappa_T^q(Q^2)$ as
obtained in the $\chi$QSM with the parameters fixed as described in
the previous Section. We first discuss the moments
$\kappa_T^q=\kappa_T^q(Q^2=0)$ and then the form factors up to a
momentum transfer of $1\,\textrm{GeV}^2$.

The anomalous tensor magnetic moment consists of the two contributions 
\begin{equation}
\kappa^q_T=-H^{*q}_T(0)-\delta q \,\,\,,
\end{equation}
where $\delta q$ is the tensor charge. The tensor charge was investigated in
the present framework in Ref.~\cite{Ledwig:2010tu} while the quantity $H^{*q}_T(0)$ is a linear combination of the
form factors of the tensor current as described in the second Section of this
work.

We will first discuss the contribution coming from the quantity
$H^*_T(0)$. In case of the $\chi$QSM it is customary to calculate first
the contributions $H_T^{*\chi=0,3,8}$, and to project afterwards on the quark
flavors u, d, s  by using Eq. (\ref{eq:flavor tensor}).
\begin{table}[ht]
\caption{\label{tab:HRMdep} The results of
  $H_{T}^{*\chi=0,3,8}(0)$ with the constituent quark mass $M$ varied
  from $400$ MeV to $450$ MeV.} 
\begin{center}

\begin{tabular}{c|rrr}\hline\hline
$M$ [MeV] & $400$ & $420$ & $450$  \\ \hline 
$H_{T}^{*0}(0)$ & $-6.29$ & $-6.15$ & $-5.76$ \\ 
$H_{T}^{*3}(0)$ & $-3.07$ & $-3.13$ & $-3.23$ \\
$H_{T}^{*8}(0)$ & $-3.38$ & $-3.54$ & $-3.72$ \\ 
\hline
\hline 
\end{tabular}
\end{center}
\end{table}
Table~\ref{tab:HRMdep} lists the results of
$H_T^{*\chi=0,3,8}$ with the constituent quark mass, the only free parameter in the $\chi$QSM,  varied from
$M=400$ MeV to $M=450$ MeV. We find that their dependence on $M$ is
rather mild, i.e. they are changed by about $3-6\, \%$ from the preferred value
with $M=420$ MeV, in line with the previous form factor calculations in the model.

In Table~\ref{tab:HR}, we list respectively each contribution of the
valence and Dirac sea quark parts, and of the SU(3) symmetric and
breaking cases for $H_T^{*\chi=0,3,8}(0)$ with $M=420$ MeV. Though the
Dirac sea contributions to the tensor charges are very
tiny~\cite{Ledwig:2010tu}, they have sizeable effects on $H_T^{*3}$ and
$H_T^{*8}$. It is again negligibly small to $H_T^{*0}$. In particular,
they contribute to $H_T^{*3,8}$ by about $40\,\%$. The effects of SU(3) 
symmetry breaking on $H_T^{*0}$ and $H_T^{*3}$, listed in the fourth
column, are small but are noticeable on $H_T^{*8}$ by about
$30\,\%$.  We want to note that the quantity $-H^*_T(0)$ is related to
the GPD $G_T(x,0,0)$ discussed in
Ref. \cite{Wakamatsu:2008}. Reference \cite{Wakamatsu:2008} does not give
integrated numbers for the charges of $G_T(x,0,0)$. However, by comparing our
numbers for the individual contributions of the SU(2) iso-scalar and iso-vector
charges listed in Table~\ref{tab:HR} with the figures given in
Ref. \cite{Wakamatsu:2008}, we find a qualitative agreement. 

\begin{table}[ht]
\caption{\label{tab:HR} The results for the singlet and non-singlet parts $H_{T}^{*\chi=0,3,8}(0)$. In the second and third columns, the 
contributions of the SU(3) valence and sea parts are listed.  
In the fourth and fifth columns we list the SU(3) symmetry breaking
contributions and the final vules, respectively. 
In the last three columns we list the SU(2) valence, sea and complete values
for the iso-singlet and iso-vector cases, $H_{T}^{*\chi=0,3}(0)$.
All results are produced with the constituent quark mass of $M=420$ MeV. } 
\begin{center}
\begin{tabular}{c|rrrr||rrr}\hline\hline
\multicolumn{1}{c|}{\phantom{111}}&\multicolumn{4}{c|}{SU(3)} & \multicolumn{3}{|c}{SU(2)} \\ 
$M=420$ MeV& Valence  & Sea & $\delta m_{\mathrm{s}}^{1}$ & Total & Valence &
Sea & Total \tabularnewline \hline 
$H_{T}^{*0}(0)$ & $-6.26$ & $-0.05$  & $0.16$ & $-6.15$&$-6.26$ &$-0.05$ &$-6.31$ \tabularnewline 
$H_{T}^{*3}(0)$ & $-1.98$ & $-1.15$ & $0.01$ &$-3.13$&$-2.03$  &$-1.30$ &$-3.33$ \tabularnewline 
$H_{T}^{*8}(0)$ & $-3.26$ & $-1.50$ & $1.22$ &$-3.54$&$--$ &$--$ &$--$\tabularnewline 
\hline
\hline 
\end{tabular}
\end{center}
\end{table}

\begin{table}[ht]

\caption{\label{tab:HstarResults} The results for the
flavor-decomposed $H_{T}^{*q}(0)$ and those of the tensor charges
$\delta q = H_{T}^{q}(0)$ Ref.~\cite{Ledwig:2010tu}. In the second and third columns, the 
contributions of the valence and sea parts are listed, respectively.  
In the fourth and fifth columns we list the SU(3) symmetry breaking
contributions and the total SU(3) values. 
For the strange form factor $H_{T}^{*s}$, we present the range of its
values with $M$ varied from 400 MeV to 450 MeV. In the last three columns we
list the corresponding SU(2) values.
}
\begin{center}

\begin{tabular}{c|cccc||ccc} \hline\hline

\multicolumn{1}{c|}{\phantom{111}}&\multicolumn{4}{c|}{SU(3)}&\multicolumn{3}{|c}{SU(2)} \\ 
 & Valence & Sea &  $\delta m_{\mathrm{s}}^{1}$ &  Total & Valence & Sea & Total  \tabularnewline  \hline 
$H_{T}^{*u}(0)$ & $-4.02$ & $-1.03$ & $0.41$ & $-4.64$&$-4.15$&$-0.67$&$-4.82$\tabularnewline 
$H_{T}^{*d}(0)$ & $-2.04$ & $0.13$ & $0.40$ & $-1.51$&$-2.11$&$0.62$&$-1.50$\tabularnewline 
$H_{T}^{*s}(0)$ & $-0.2\sim-0.3 $ & $0.7\sim 1.0$ &$-0.7\sim -0.6$ & $-0.2\sim0.2$&$--$&$--$&$--$   \tabularnewline 
\hline 
$\delta u$ & $1.01$ & $0.07$ & $-0.01$ &$1.08$&$1.02$&$0.07$&$1.09$\tabularnewline 
$\delta d$ & $-0.28$ & $-0.03$ & $-0.01$ &$-0.32$&$-0.30$&$-0.04$&$-0.34$\tabularnewline 
$\delta s$ & $-0.02$ & $\approx0$ & $0.01$ &$-0.01$&$--$&$--$&$--$\tabularnewline \hline\hline
\end{tabular}
\end{center}
\end{table}

We will now turn to the flavor-decomposed contributions of the charges
$H_T^{*q}(0)$ and tensor charges $\delta q$. 
In Table~\ref{tab:HstarResults} we list the two parts $H_T^{*q}(0)$ and
  $\delta q$ contributing to the anomalous tensor magnetic moment $\kappa_T^q$ separately. The tensor charges $\delta
    q$ were studied in Ref.~\cite{Ledwig:2010tu} within exactly the same
    framework. The results for for the SU(3) and SU(2) versions of the
$\chi$QSM are presented in Table~\ref{tab:HstarResults}. Comparing the results of these two versions, we see that they give
nearly the same total values for the up and down quarks, $H_T^{*u}(0)$ and
$H_T^{*d}(0)$. However, the individual decompositions are different. While the
valence quark contribution is comparable in both versions, the Dirac-sea
contributions show different features. In the case of
SU(2), the Dirac-sea contribution nearly vanishes for the iso-scalar case. This
has the consequence that the Dirac-sea contributions to up and down quarks come almost
completely from the iso-vector case and therefore yield the same absolute
value but with opposite sign. This can also be seen in the SU(2)
$\chi$QSM work \cite{Wakamatsu:2008}. In the case of the SU(3) version, it is
the {\it sum} of the Dirac-sea and the strange quark mass correction which nearly gives the same
contribution in absolute value to the valence part and is comparable to the SU(2)
Dirac-sea component.\\   

We now come to the strange form factor $H_T^{*s}$. We find that the sea quark contribution is
substantially larger than the valence one. This can be understood
qualitatively from Eq. (\ref{eq:38CQSMdensity}) in which we have $r^3$ in the integral. This
factor amplifies the long-range tail of the densities when we carry
out the integral.  A similar result can be found in the neutron
electric radius~\cite{Kim:eleff,Christov:1995vm} in which the Dirac
sea part is almost the same order of the valence one. However, the
dependence of $H_T^{*s}$ on the constituent quark mass $M$ is also
rather sensitive. Thus, we give the results for $H_T^{*s}$ in the
range from $M=400$ to $450$ MeV  and will regard this as our theoretical
uncertainty.   

The effects of SU(3) symmetry breaking are mild but
nonnegligible on $H_T^{*u}$ and $H_T^{*d}$. They are
oppositely polarized to the valence part, so that they reduce
the magnitudes of $H_T^{*u}$ and $H_T^{*d}$ by almost $10\,\%$ and $20\,\%$, respectively. The contribution of SU(3)
symmetry breaking is noticeably large in the strange form factor 
$H_T^{*s}$.  Moreover, it depends sensitively on the constituent quark
mass $M$. In the case of $M=420$ MeV, it cancels out the SU(3)
symmetric part, so that $H_T^{*s}(0)$ almost vanishes.
We want to stress that the form factors $H_T^{\chi=0,3,8}$ are insensitive to
the constituent quark mass $M$. They are generally 
changed by about $5\,\%$ with $M$ varied from 400 MeV to 450
MeV. However, $H_T^{*s}$ shows strong dependence on $M$.   
Because of this behavior of $H_T^{*s}(0)$ we conclude that this quantity is at
the numerical limit of the $\chi$QSM. However, what we do infer from the given
values for $H_T^{*s}(0)$ is the fact that the individual parts always destructively
interfere for a given $M$. Hence, the $\chi$QSM yields a small $H_T^{*s}(0)$ and
together with the small value of $\delta s$ the $\chi$QSM predicts a small
strange anomalous tensor magnetic moment $\kappa_T^s$ of the proton.

\begin{table}[ht]
\caption{\label{tab:ATMM}The results for the flavor-decomposed
anomalous tensor magnetic moments $\kappa_{T}^{q}$. In the second
and third columns, the contributions of the valence and sea parts
are listed, respectively.  In the fourth and fifth columns we list
the SU(3) symmetry breaking contributions and the total value for the SU(3)
$\chi$QSM. For the strange form factor $\kappa_{T}^{s}$, we present
its results in the range with $M$ varied from 400 MeV to
450 MeV. In the last three columns we list the corresponding values for the
SU(2) $\chi$QSM version.}  
\begin{center}

\begin{tabular}{c|cccc||ccc} \hline\hline
\multicolumn{1}{c|}{\phantom{111}}&\multicolumn{4}{c|}{SU(3)}&\multicolumn{3}{|c}{SU(2)} \\ 
Proton & Valence & Sea & $\delta m_{\mathrm{s}}^{1}$ & Total & Valence & Sea &
Total \tabularnewline  \hline 
$\kappa_{T}^{u}$ & $3.01$ & $0.96$ & $-0.40$ &$3.56$&$3.13$&$0.60$&$3.72$\tabularnewline 
$\kappa_{T}^{d}$ & $2.32$ & $-0.10$ & $-0.39$ &$1.83$&$2.42$&$-0.58$&$1.83$\tabularnewline 
$\kappa_{T}^{s}$ & $0.2\sim0.3$ & $-0.7\sim-1.0$ & $0.7\sim0.6$ & $0.2\sim-0.2$&$--$&$--$&$--$\tabularnewline 
\hline\hline
\end{tabular}

\end{center}
\end{table}

The results for the anomalous tensor magnetic moments $\kappa_T^q$ can easily
be read off from Table~\ref{tab:HstarResults}, since they are given by the combination
$\kappa_T^q =-H_T^{*q}-\delta q,$.
The final results are listed in Table~\ref{tab:ATMM} for the SU(3) and SU(2)
$\chi$QSM versions. In both versions we obtain nearly the same values for the
up and down contributions $\kappa_T^u$ and $\kappa_T^d$. As already seen for the quantity
$H_T^{*q}(0)$, the valence parts for SU(3) and SU(2) are comparable, while
residual contributions are decomposed differently. Again it is the sum of the
Dirac-sea and strange quark mass corrections of the SU(3) version which is comparable to the
Dirac-sea contribution of the SU(2) version. In the case of the strange anomalous
magnetic moment $\kappa_{T}^{s}$ the numerical uncertainty observed in
$H_T^{*s}$ is carried over also to $\kappa_{T}^{s}$. However, for all numerical settings we obtain the
common feature that the individual contributions to $\kappa_T^s$ destructively
interfere.
The final results is given as the interval $\kappa_T^s = 0.2 \sim -0.2 $
corresponding to the change of the $\chi$QSM parameter $M$ between $400$ MeV and
$450$ MeV. In total, because of this numerical sensitivity on $M$, we consider
$\kappa_T^s$ as to be at the numerical limit of the $\chi$QSM. Nevertheless, from
the destructive interference observed comonly in given numerical setups,
we can conclude that the $\chi$QSM
predicts a small, nearly vanishing $\kappa_T^s$.

\begin{table}[ht]
\caption{\label{tab:COMP}Comparison of the results for the anomalous tensor magnetic moments
$\kappa_T^q$ with other works. In the first two columns, the final results of the
  present $\chi$QSM work in the SU(3) and SU(2) version are listed.  In the second and third columns, we list the corresponding
results of the SU(2) $\chi$QSM~\cite{Wakamatsu:2008} and those of the
lattice calculation~\cite{Hagler:2008}. The lattice calculation was done at a
scale of $4\,\textrm{GeV}^2$ where we give in parantheses the corresponging
value at the $\chi$QSM scale of $0.36\,\textrm{GeV}^2$. The last column lists
the results of the light-front constituent quark model (LFCQM)~\cite{Paspquini:2007}.  We also
list the renormalization-independent ratio
$\kappa_{T}^{u}/\kappa_{T}^{d}$ in the last row. For the strange form
factor $\kappa_{T}^{s}$, we present
its results in the range with $M$ varied from 400 MeV to
450 MeV.}

\begin{tabular}{c|cc|ccc}\hline\hline
 &Present work SU(3) & Present work SU(2)&  $\chi$QSM SU(2) \cite{Wakamatsu:2008}  &
 Lattice~\cite{Hagler:2008}   &
 LFCQM~\cite{Paspquini:2007}\tabularnewline  
\hline
$\kappa_{T}^{u}$ & $3.56$&$3.72$ & $3.47$ & $3.00\,\,(3.70)$ &$3.98$\tabularnewline 
$\kappa_{T}^{d}$ & $1.83$&$1.83$ & $2.60$ & $1.90\,\,(2.35)$ &$2.60$\tabularnewline 
$\kappa_{T}^{s}$ & $0.2\sim-0.2$ & $ $
& \tabularnewline 
\hline
$\kappa_{T}^{u}/\kappa_{T}^{d}$ & $1.95$&$2.02$ & $1.33$ & $1.58$ &
$1.53$\tabularnewline \hline\hline
\end{tabular}
\end{table}

We now compare in Table~\ref{tab:COMP} our results for the anomalous tensor 
magnetic moment with those of other 
works~\cite{Wakamatsu:2008,Paspquini:2005,Hagler:2008}.
Wakamatsu~\cite{Wakamatsu:2008} investigated $\kappa_T^q$ within the SU(2) 
$\chi$QSM and Ref.~\cite{Hagler:2008} computed this quantity on the lattice. 
In Ref.~\cite{Paspquini:2007} the anomalous tensor magnetic moment
was studied in SU(6) symmetric light-front constituent quark
models. We present also the renormalization
scale-invariant ratio $\kappa_{T}^{u}/\kappa_{T}^{d}$.   

The main differences of the given approaches are as follows. 
In the case of the SU(2) $\chi$QSM \cite{Wakamatsu:2008} a constituent quark mass
of $M=375$ MeV was used together with the Pauli-Villars regularization. The
moments were obtained by first calculating the corresponding GPDs and
integrating over the fraction parameter $x$ afterwards. The present work uses
$M=420$ MeV, the proper-time regularization and calculates the corresponding
generalized form factor at $Q^2=0$. Compared to Ref. \cite{Wakamatsu:2008} the
present SU(2) results differ for $\kappa_T^u$ by $7$ \% and for $\kappa_T^d$
by $30$ \%. Since both SU(2) works use different approaches with
different numerical methods and soliton profiles, a deviation of $30$ \% is
acceptable for the given model.
The lattice results \cite{Hagler:2008} are calculated for the finite momentum
transfers of $Q^2=0.4\sim2.5 \,\textrm{GeV}^2$, a pion mass of $m_\pi=600$
MeV and a lattice spacing of $a\approx0.08$ fm at a renormalization scale of
$\mu^2=4\, \textrm{GeV}^2$. The moments are obtained by
extrapolating to $Q^2=0$ with a p-pole ansatz and linearly extrapolated to the physical
pion mass. The work of Ref.~\cite{Paspquini:2007} uses an SU(6)
symmetric light-front consitutent quark model on the valence
quark level. In total $\kappa_T^u$ seem to agree each other in all approaches while
the absolute value for $\kappa_T^d$ in the present work is lower than most of
the others. However, the overall tendency $\kappa_T^u > \kappa_T^d>0$ is seen
in all works. This work gives additionally the value of $\kappa_T^s \approx
0$.

We turn now to the discussion of the full form factors for a finite momentum transfer.

\begin{figure}[ht]
\caption{\label{fig:KapapFF} (Color online) The results for the SU(3)
flavor-decomposed anomalous tensor magnetic form factors. The solid
curve draws the up anomalous tensor magnetic form factor, whereas the
dashed and short-dashed ones depict the down and strange ones, respectively. The
constituent quark mass $M=420$ MeV is used.} 

\includegraphics[scale=0.7]{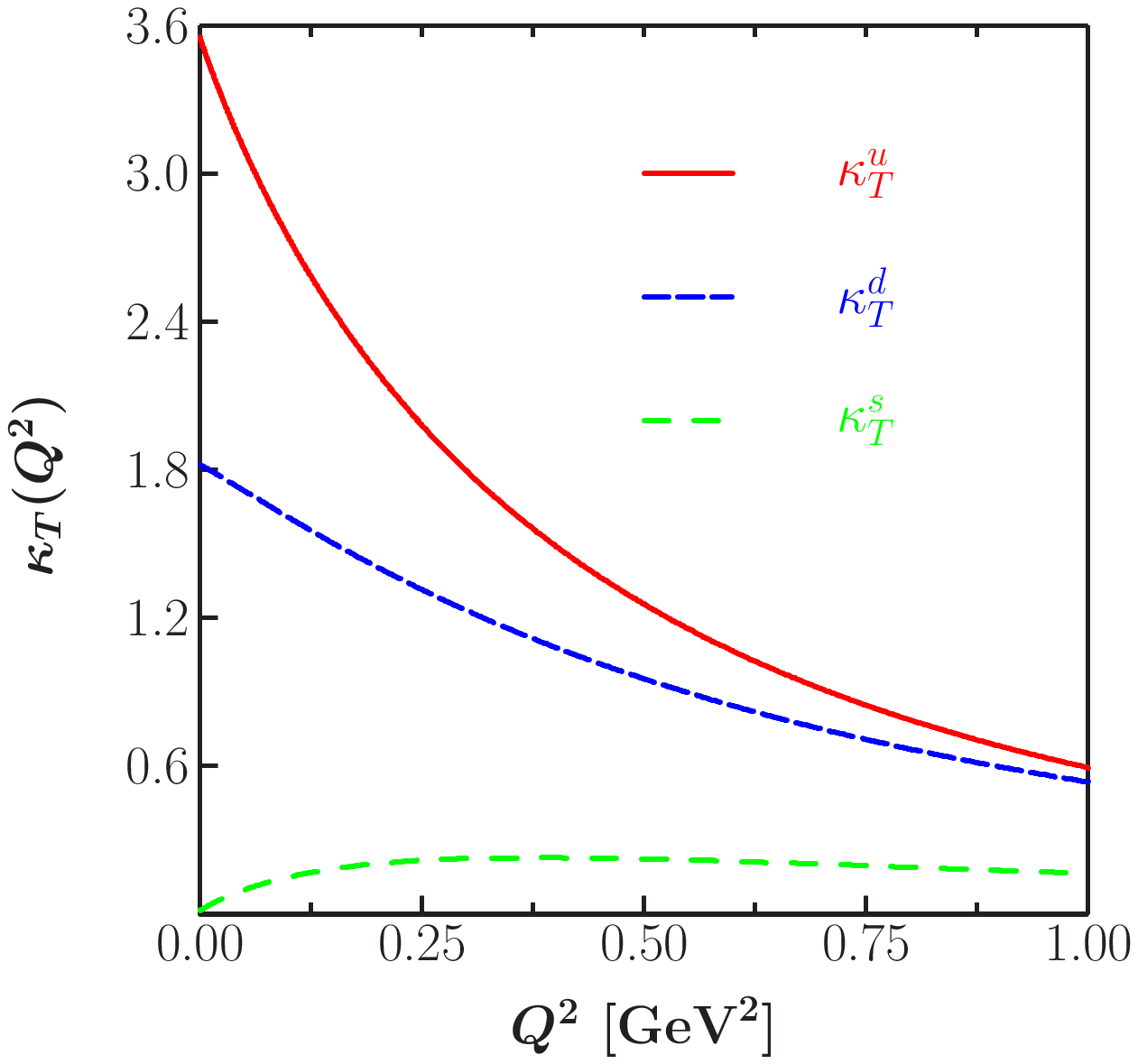}

\end{figure}

The flavor-decomposed anomalous tensor magnetic form factors
$\kappa_{T}^{q}(Q^{2})$ are drawn in Fig.~\ref{fig:KapapFF} for the constituent
quark mass of $M=420$ MeV and for the momentum transfer up to $Q^2=1\, \textrm{GeV}^2$. The up
and down form factors fall off as $Q^2$ increases, while the strange
one starts to increase slowly and then gets lessened mildly from
around $0.25$ $\textrm{GeV}^2$ as $Q^2$ grows. The feature of the strange form
factor seems very similar to the neutron electric form factor.  

\begin{figure}[ht]
\caption{\label{fig:Kapalat} (Color on line) The comparison of the
up and down anomalous tensor magnetic form factors with lattice results. The solid curves draw the results of the present
work with $M=420$ MeV, whereas the dashed ones depict the lattice
results~\cite{Hagler:2008}. In the left panel, the result of the up
anomalous tensor magnetic form factor is compared to that of the
lattice. The right panel is for the down form factors. The
lattice calculation was performed with $m_{\pi}=600$ MeV.} 
\includegraphics[scale=0.56]{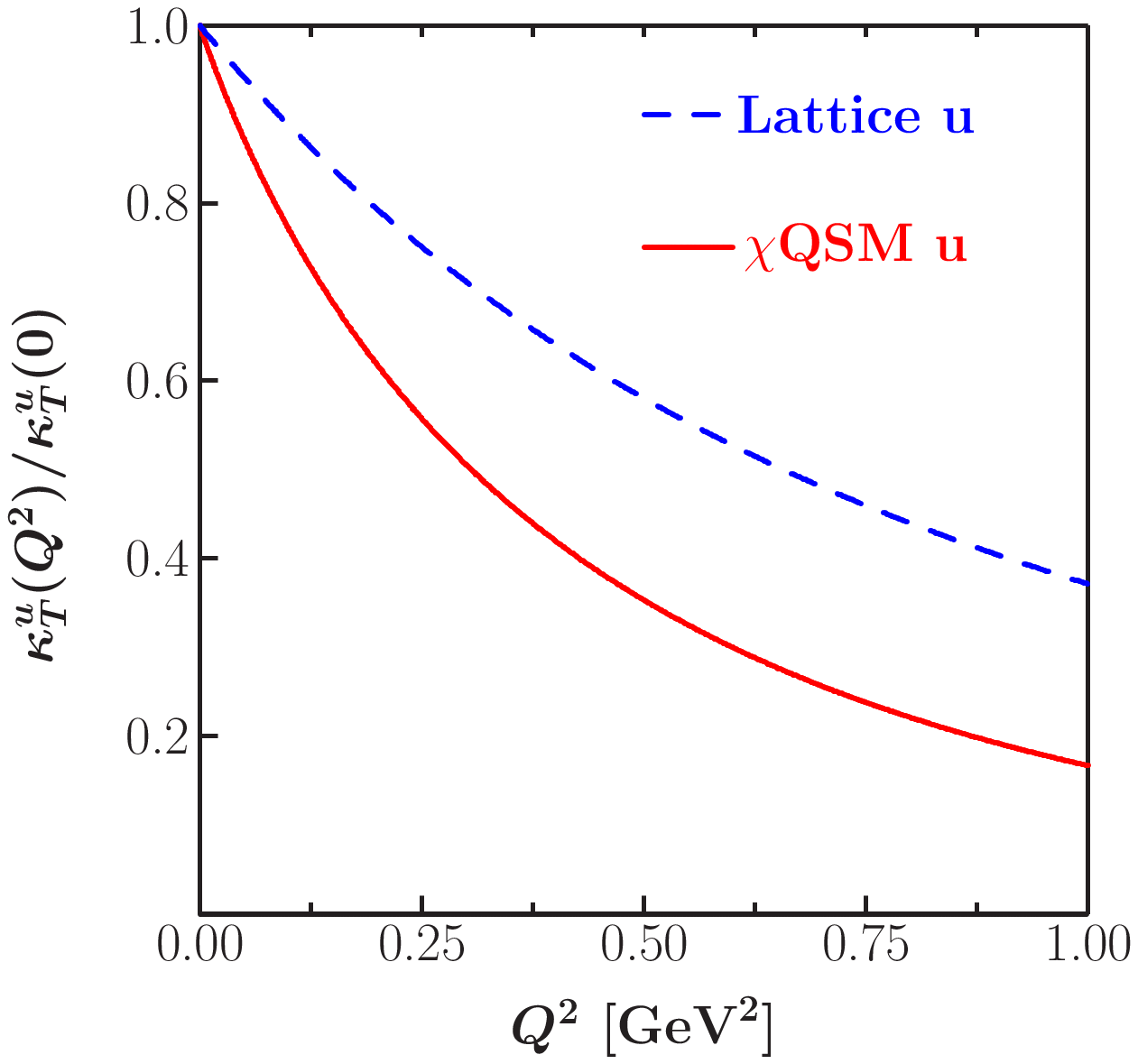}~~~~~~
\includegraphics[scale=0.56]{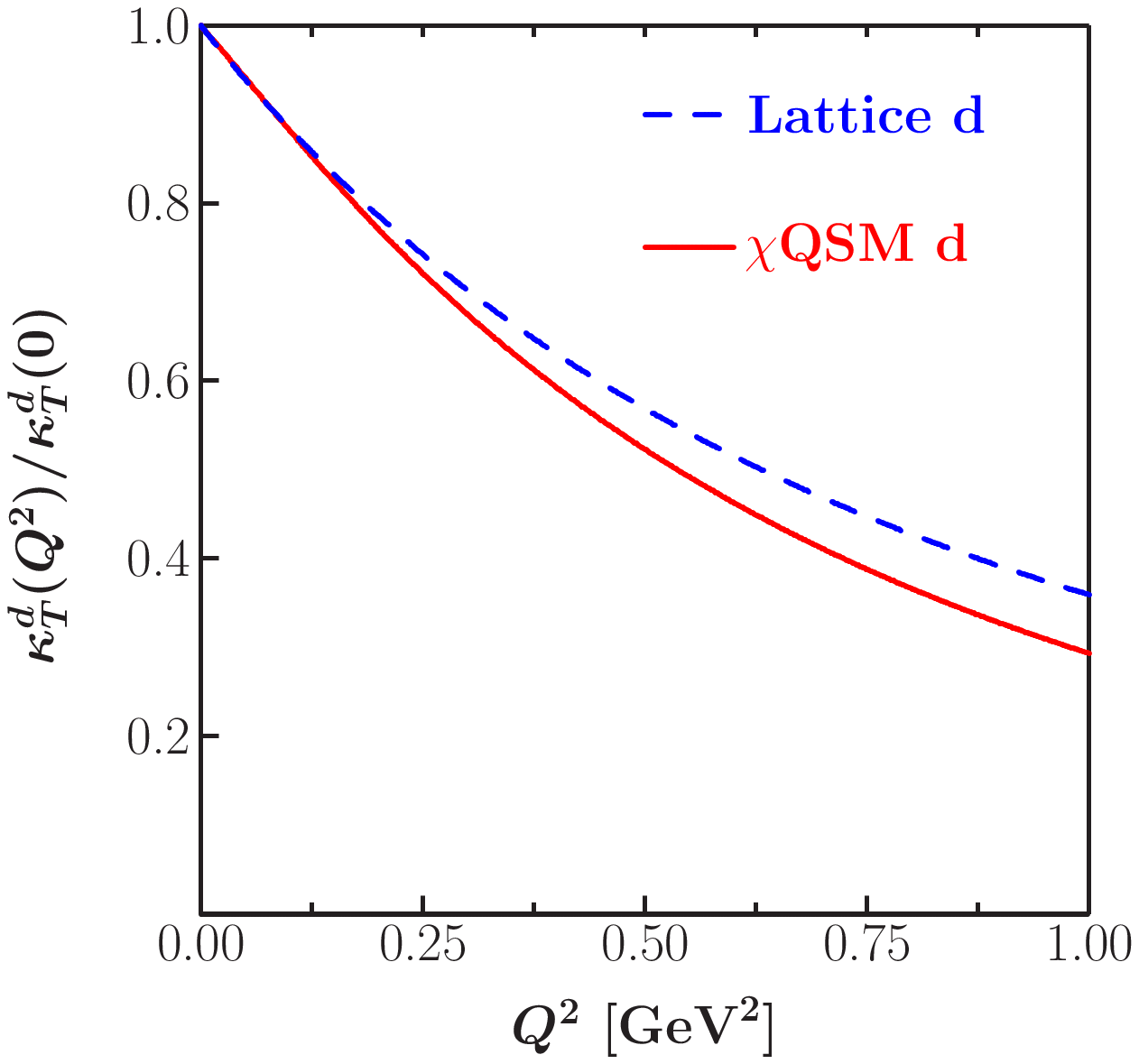}
\end{figure}

In Fig.~\ref{fig:Kapalat} we compare the present results of the up and
down anomalous tensor magnetic form factors with those of the
lattice~\cite{Hagler:2008}. In order to compare the $Q^2$ dependence
directly, we have scaled the form factors by the corresponding values
at $Q^2=0$. This has also the advantage that the renormalization scale dependence is
canceled out.
In the left panel, we first show the
present up form factor and the corresponding lattice one. The lattice result
decreases rather slowly as $Q^2$ increases, compared to the present result. This is a natural tendency of the lattice calculation because of
the large value of the pion mass. We find the similar behavior in the
case of the tensor form factor $\delta u(Q^2)$~\cite{Ledwig:2010tu}.  
In the right panel, we draw the result of the down form factor in
comparison with that of the lattice calculation.  Interestingly, the
present result shows almost a similar $Q^2$ dependence to the lattice
one.  

It is also interesting to parametrize the up and down form factors in
a simple form such as
\begin{equation}
\kappa_{T}^q(Q^{2}) \; = \;
\frac{\kappa_{T}^q(0)}{\left(1+\frac{Q^{2}}{\alpha M_\alpha^{2}}\right)^{\alpha}}\, .   
\label{eq:generalFF} 
\end{equation} 
The results for the power $\alpha$ and the pole mass $M_\alpha$ are given in
Table~\ref{tab:parameters}.
\begin{table}[ht]
\caption{\label{tab:parameters} The parameters for
  Eq.(\ref{eq:generalFF}) fitted to the present results of the
  anomalous  tensor magnetic form form factors.}
\begin{tabular}{c|ccc||c|ccc} \hline\hline 
 & $\kappa_{T}^\chi(0)$ & $\alpha$& $M_\alpha\,[\mathrm{GeV}]$&& $\kappa_{T}^q(0)$ & $\alpha$ & $M_\alpha\,[\mathrm{GeV}]$\tabularnewline
\hline
$\kappa_{T}^{0}(Q^{2})$ & $5.39$ & $2.82$ & $0.74$&$\kappa_{T}^{u}(Q^{2})$ & $3.55$ & $2.28$ & $0.61$\tabularnewline
$\kappa_{T}^{3}(Q^{2})$ & $1.73$ & $5.33$ & $0.49$&$\kappa_{T}^{d}(Q^{2})$ & $1.82$ & $8.12$ & $0.86$ \tabularnewline
$\kappa_{T}^{8}(Q^{2})$ & $3.09$ & $2.30$ & $0.59$& \tabularnewline
\hline\hline
\end{tabular}
\end{table}
Actually, G\"ockeler et al.~\cite{Hagler:2008} have used the ansatz
given in Eq.(\ref{eq:generalFF}) to fit the lattice data. The lattice
results of $M_\alpha$ for $\kappa_T^u$ and $\kappa_T^d$ are given, respectively, as
\begin{equation}
  \label{eq:latticefit}
M_\alpha^{(u)} \;=\; 0.907\,\mathrm{GeV},\;\;\;
M_\alpha^{(d)} \;=\; 0.889\,\mathrm{GeV}.
\end{equation}
The power $\alpha=2.5$ was used for all the lattice data. The
Fig.~\ref{fig:Kapalat} can be reproduced with the above given data.




\section{Summary and outlook}

In the present work, we aimed at investigating the anomalous tensor
magnetic moments $\kappa_T^q$ and its corresponding form factors of the proton within the framework of the
self-consistent SU(3) and SU(2) chiral quark-soliton model ($\chi$QSM).  
Generally, the anomalous tensor magnetic moments are given by two
contributions: $\kappa_T^q=-H^{*q}_T-\delta q$ where $\delta q$ is the tensor
charge and $H^{*q}_T$ is a certain linear combination of the form factors
appearing in the nucleon matrix element of the tensor current.
We first computed the quantities $H^{*q}_T$ in the SU(3) and SU(2) version of
the $\chi$QSM. We saw that, in contrast to the tensor charges $\delta q$, the
quantities $H^{*u,d,s}_T$ have in the present framework noticeable
contributions from the Dirac-sea and linear strange quark mass
corrections. Interestingly, in the case of the SU(3) version these two
contributions arrange each other in such a way that their combined SU(3) value
is comparable to the SU(2) Dirac-sea value. As a whole, since the valence quark
contribution in both versions is nearly the same,  we obtain therefore for the
components $H^{*u,d}_T$ in both versions consistent (comparable) values. For the strange quantity $H^{*s}_T$ we see a
rather sensitive behavior with respect to the constituent quark mass, the only
free parameter in the $\chi$QSM. We regard therefore this quantity to be at
the numerical limit of accuracy of the $\chi$QSM. However, we observed for every given
numerical setup that the individual parts of $H^{*s}_T$ tend to cancel
each other. We infer from this that the $\chi$QSM predicts a small $H^{*s}_T$.

The tensor charges $\delta q$ were investigated with the present framework in 
Ref.~\cite{Ledwig:2010tu}. Combining the result of that work with the quanties
$H^{*q}_T$ of the present one, we were able to predict the anomalous
 tensor magnetic moments $\kappa_T^{u,d,s}$. The 
values of $\kappa_T^{u,d}$ turned out to be positive with the SU(3) values
$\kappa_T^u=3.56$ and $\kappa_T^d=1.83$ while those of
$\kappa_T^s$ were given in the range of $\kappa_T^s=0.2\sim-0.2$. This
numerical uncertainty for $\kappa_T^s$ is carried over from $H_T^{*s}$ as well
as the tendency that the individual parts of $\kappa_T^s$ are destructively
interfereing for any given numerical setup. Accordingly to $H^{*s}_T$, we
infer that $\kappa_T^s$ is at the numerical limit of accuracy of the $\chi$QSM but that a
small, nearly vanishing $\kappa_T^s$ is predicted. The SU(2) $\chi$QSM values
of $\kappa_T^{u,d}$ are: $\kappa_T^u=3.72$ and $\kappa_T^d=1.83$ which are comparable to the SU(3) counterparts.
We compared the present results of $\kappa_T^{u,d}$ with those of other
works.  They are in qualitative agreement with each other with the hierachy:
$4>\kappa_T^u>\kappa_T^d>0$ for a renormalization scale below
$4\,\textrm{GeV}^2$.
  
We also investigated the anomalous tensor magnetic form factors
$\kappa_{T}^q(Q^{2})$ of the proton up to a momentum transfer of
$Q^{2}=1\,\mathrm{GeV}^{2}$.  While the up and down form factors
$\kappa_T^u$ and $\kappa_T^d$ fall off as $Q^2$ increases,
the strange form factor $\kappa_T^s$ starts to grow slowly and then
gets lessened mildly from around $0.25$ GeV as $Q^2$ increases. This
$Q^2$ dependence of the strange form factor seems very similar to the
neutron electric form factor. We compared the results of the up and
down form factors with those of the lattice and found that while for
$\kappa_T^u$ the lattice result falls off more slowly than the present
one, the present result for $\kappa_T^d$ is pretty much similar to the
lattice one.   
By combining the calculation of the electromagnetic form factors with the
present results and those of the tensor form
factors~\cite{Ledwig:2010tu}, we will be able to get access to the first
moment of the transverse quark spin density which yields information on
the correlation of transverse coordinate and spin degrees of freedom
that are related to the Sivers and Boer-Mulders
functions~\cite{Burkardt:2005,Burkardt:2004a}. The corresponding
investigation is under way. 

\section*{Acknowledgments}
The authors are grateful to P. H\"agler, B. Pasquini, P. Schweitzer,
and M. Vanderhaeghen for valuable discussions and critical
comments. T.L. was supported by the Research Centre
``Elementarkr\"afte und Mathematische Grundlagen'' at 
the Johannes Gutenberg University Mainz. The present work is also supported
by Basic Science Research Program through the National Research
Foundation of Korea (NRF) funded by the Ministry of Education, Science
and Technology (Grant No. 2009-0073101).

\begin{appendix}

\section{Densities}
In this Appendix, we provide the densities for the tensor form factors 
given Eq.~(\ref{eq:38CQSMdensity}), which consist of
$\mathcal{Q}_{T0}(r),\cdots,\mathcal{M}_{T2}(r)$. In the following,
the sums run freely over all single-quark levels including valence
ones except that the sum over $m_0$ is constrained to negative-energy
levels: 
\begin{eqnarray}
\frac{1}{N_{c}}\mathcal{Q}_{T0}(r) & = & \langle \mathrm{val}|r\rangle
i\gamma_4\{\frac{z_{1}}{r}\otimes\sigma_{1}\}_{0}\gamma^{5}\langle
r|\mathrm{val}\rangle \cr
&+& \sum_{n}
\sqrt{2G_{n}+1}\mathcal{R}_{1}(\varepsilon_{n})\langle n|r\rangle 
i\gamma_4\{\frac{z_{1}}{r}\otimes\sigma_{1}\}_{0}\gamma^{5}\langle
r|n\rangle,\\ 
\frac{1}{N_{c}}\mathcal{X}_{T1}(r) & = &
\sum_{\varepsilon_{n}\neq\varepsilon_{v}}
\frac{1}{\varepsilon_{n}-\varepsilon_{v}}(-)^{G_{n}}\langle 
\mathrm{val}|\tau_{1}|n\rangle\langle n|r\rangle 
i\gamma_4\{\{\frac{z_{1}}{r}\otimes\sigma_{1}\}_{0}
\otimes\tau_{1}\}_{1} \gamma^{5}\langle r|\mathrm{val}\rangle\cr 
 & + &
 \frac{1}{2}\sum_{n,m} \mathcal{R}_{5}(\varepsilon_{n},
 \varepsilon_{m}) (-)^{G^{m}-G_{n}}\langle
 n|\tau_{1}|m\rangle\langle m|r\rangle
 i\gamma_4\{\{\frac{z_{1}}{r}\otimes\sigma_{1}\}_{0} \otimes
 \tau_{1}\}_{1} \gamma^{5}\langle
 r|n\rangle,\\  
\frac{1}{N_{c}}\mathcal{X}_{T2}(r) & = &
\sum_{n^{0}} \frac{1}{\varepsilon_{n^{0}} -
  \varepsilon_{v}} \sqrt{2G_{n}+1}\left[\langle
  \mathrm{val}|n^{0}\rangle\langle \mathrm{val}|r\rangle
  i\gamma_4\{\frac{z_{1}}{r}\otimes\sigma_{1}\}_{0}\gamma^{5}\langle
  r|n^{0}\rangle\right]\cr 
 & + & \sum_{n,m^0} \mathcal{R}_{5}(\varepsilon_{n},
 \varepsilon_{m^0})\sqrt{2G_{n}+1}\left[\langle 
   n|m^{0}\rangle\langle n| r\rangle
   i\gamma_4\{\frac{z_{1}}{r}\otimes\sigma_{1}\}_{0}\gamma^{5}\langle
   r|m^{0}\rangle\right],\\ 
\frac{1}{N_{c}}\mathcal{M}_{T0}(r) & = &
\sum_{\varepsilon_{n}\neq\varepsilon_{v}}
\frac{1}{\varepsilon_{n}-\varepsilon_{v}}\langle 
\mathrm{val}|r\rangle
i\gamma_4\{\frac{z_{1}}{r}\otimes\sigma_{1}\}_{0}\gamma^{5}\langle
r|n\rangle\langle n|\gamma_4|\mathrm{val}\rangle\cr
 & - & \frac{1}{2}\sum_{n,m}\sqrt{2G_{n}+1}\langle
 n|\gamma_4|m\rangle\langle m|r\rangle
 i\gamma_4\{\frac{z_{1}}{r}\otimes\sigma_{1}\}_{0}\gamma^{5}\langle
 r|n\rangle\mathcal{R}_{2}(\varepsilon_{n},\varepsilon_{m}), \\ 
\frac{1}{N_{c}}\mathcal{M}_{T1}(r) & = &
\sum_{\varepsilon_{n}\neq\varepsilon_{v}}
\frac{(-)^{G_{n}}}{\varepsilon_{n}-\varepsilon_{v}} \langle \mathrm{val}|r\rangle  
i\gamma_4\{\{\frac{z_{1}}{r}\otimes\sigma_{1}\}_{0}\otimes\tau_{1}\}_{1}
\gamma^{5} \langle r|n\rangle\langle n|\gamma_4\tau_{1}|\mathrm{val}\rangle \cr
 & - &  \frac{1}{2}\sum_{n,m}(-)^{G_{m}-G_{n}}\langle
 n|\gamma_4\tau_{1}|m\rangle\langle m|r\rangle
 i\gamma_4\{\{\frac{z_{1}}{r}\otimes\sigma_{1}\}_{0}\otimes\tau_{1}\}_{1}
 \gamma^{5}\langle
 r|n\rangle\mathcal{R}_{2}(\varepsilon_{n},\varepsilon_{m}),\\  
\frac{1}{N_{c}}\mathcal{M}_{T2}(r) & = &
\sum_{n^{0}} \frac{1}{\varepsilon_{n^{0}}-\varepsilon_{v}}\langle
\mathrm{val}|r\rangle
i\gamma_4\{\frac{z_{1}}{r}\otimes\sigma_{1}\}_{0}\gamma^{5}\langle
r|n^{0}\rangle\langle n^{0}|\gamma_4|\mathrm{val}\rangle\cr 
 & - & \sum_{n,m^0}\sqrt{2G_{n}+1}\langle m^{0}|r\rangle
 i\gamma_4\{\frac{z_{1}}{r}\otimes\sigma_{1}\}_{0} \gamma^{5}\langle
 r|n\rangle\langle
 n|\gamma_4|m^{0}\rangle\mathcal{R}_{2}(\varepsilon_{n},\varepsilon_{m^0}),
\end{eqnarray} 
\end{appendix}
where the regularization functions are given by:
\begin{eqnarray}
\mathcal{R}_{1}(\varepsilon_{n}) & = & -\frac{1}{2\sqrt{\pi}}\varepsilon_{n}\int_{1/\Lambda^{2}}^{\infty}\frac{du}{\sqrt{u}}e^{-u\varepsilon_{n}^{2}}\,\,,\\
\mathcal{R}_{2}(\varepsilon_{n},\varepsilon_{m}) & = & \int_{1/\Lambda^{2}}^{\infty}du\frac{1}{2\sqrt{\pi u}}\frac{\varepsilon_{m}e^{-u\varepsilon_{m}^{2}}-\varepsilon_{n}e^{-u\varepsilon_{n}^{2}}}{\varepsilon_{n}-\varepsilon_{m}}\,\,,\\
\mathcal{R}_{5}(\varepsilon_{n},\varepsilon_{m}) & = & \frac{1}{2}\frac{\textrm{sign}\varepsilon_{n}-\textrm{sign}\varepsilon_{m}}{\varepsilon_{n}-\varepsilon_{m}}\,\,.
\end{eqnarray}

\end{document}